\documentclass[showpacs,preprint]{revtex4}
\usepackage{graphicx}

\usepackage{amssymb}
\usepackage{amsmath}

\begin{document}

\title{Cavitation and bubble collapse in hot asymmetric nuclear matter.}
\author{V.M. Kolomietz \footnote{%
Permanent address: Institute for Nuclear Research, Prospect Nauki
47,\ 03680 Kiev, Ukraine \newline e-mail: vkolom@kinr.kiev.ua }}
\address{Physik-Department TU M\"{u}nchen, 85747 Garching, Germany}

\begin{abstract}
The dynamics of embryonic bubbles in overheated, viscous and non-Markovian
nuclear matter is studied. We show that the memory and the Fermi surface
distortions significantly affect the hinderance of bubble collapse and
determine a characteristic oscillations of the bubble radius. These
oscillations occur due to the additional elastic force induced by the memory
integral.
\end{abstract}

\maketitle

  1. Growing or collapsing bubbles in an overheated (undercompressed)
fluid
has been a intriguing problems of physics over a long period of time \cite%
{lohse}. In nuclear physics, the bubbles may be formed in the expanded
nuclear matter in a heavy-ion reaction then providing the break-up of the
matter into fragments \cite{bond1,barz,snep}. The formation of embryonic
bubbles in hot nuclear matter due to the quantum and statistical
fluctuations has been studied earlier in Refs. \cite%
{scvo83,blin86,blin,bodo88,dosn89,sach92}. The goal of the present paper is
the analysis of the dynamic evolution and the collapse of the bubble in an
overheated nuclear matter.

In a classical liquid, the dynamics of the bubble of radius $R$ is derived
mainly from the thermodynamic potential $\Phi (R)$, which reaches a \textit{%
maximum} value for the critical radius $R=R^{\ast }$ \cite{frenk}. More
complicated bubble dynamics occurs in the case of a Fermi liquid where the
dynamic distortion of the Fermi surface produces an additional pressure
tensor \cite{ak,kota}. Moreover, the Fermi surface distortion and the
interparticle collisions lead here to the non-Markovian equations of motion
for the relevant collective variables \cite{kora01}. Below the non-Markovian
dynamics will be applied to the problem of the collapse of bubbles. The
non-Markovian equation of motion for the bubble radius $R$ will be derived
in a hot nuclear Fermi-liquid. However, the final macroscopic equations of
motion including the memory kernel can be applied to a wide number of
non-Newtonian liquids.

2. We will consider the dynamics of the bubble (cavitation) with an
arbitrary undercritical size $R\leq R^{\ast }$, where the critical radius $%
R^{\ast }$ of the bubble in an overheated liquid at the temperature $T$. The
critical radius $R^{\ast }$ depends on the overheating temperature $\Delta
T=T-T_{0}$, where $T_{0}$\ is the boiling temperature in the case of plane
geometry. For fixed values of temperature $T$\ and liquid phase pressure $P_{%
\mathrm{liq}}$\ the critical radius $R^{\ast }$ is derived from the
condition of thermodynamic equilibrium of the vapor bubble with the
surrounding liquid and it is given by \cite{ll-1,frenk}
\begin{equation}
R^{\ast }=\frac{2\sigma T_{0}}{\rho _{\mathrm{vap}}\overset{\_}{\phi }\Delta
T}.  \label{rcrit}
\end{equation}%
Here, $\sigma $\ is the surface tension coefficient, $\rho _{\mathrm{vap}}$\
is the particle density in the bubble and $\overset{\_}{\phi }$ is the
latent heat of evaporation. Considering the dynamics of the surrounding
Fermi liquid, one can reduce the collisional kinetic equation for the
phase-space distribution function to Euler-like equations for the velocity
field $\mathbf{u}(\mathbf{r},t)$ in the form (for details, see Refs. \cite%
{kota,kora01,kosh04})
\begin{equation}
m{\frac{\partial }{\partial t}}u_{\nu }+m\rho (u_{\mu }\nabla _{\mu })\
u_{\nu }+\frac{1}{\rho }\nabla _{\nu }\mathcal{P}+\nabla _{\nu }{\frac{%
\delta \mathcal{\epsilon }_{\mathrm{pot}}}{\delta \rho }}=-\frac{1}{\rho }%
\nabla _{\mu }\mathcal{P}_{\nu \mu }^{\prime }+F_{\nu ,\mathrm{ext}},
\label{e2}
\end{equation}%
where $\rho \equiv \rho (\mathbf{r},t)$\ is the particle density, $\mathcal{%
\epsilon }_{\mathrm{pot}}\equiv \mathcal{\epsilon }_{\mathrm{pot}}(\mathbf{r}%
,t)$\ is the potential energy density and $\mathcal{P\equiv P}(\mathbf{r},t)$
is the pressure caused by both the thermal and Fermi motions of nucleons.
The pressure tensor $\mathcal{P}_{\nu \mu }^{\prime }\equiv \mathcal{P}_{\nu
\mu }^{\prime }(\mathbf{r},t)$ in Eq. (\ref{e2}) is caused by the Fermi
surface distortion. In the case of the most important quadrupole distortion
of the Fermi surface, the pressure tensor $\mathcal{P}_{\nu \mu }^{\prime }$%
\ satisfies the following equation \cite{kora01}\

\begin{equation}
{\frac{\partial }{\partial t}}\mathcal{P}_{\nu \mu }^{\prime }+\mathcal{P\ }{%
\frac{\partial }{\partial t}}\Lambda _{\nu \mu }=-\frac{1}{\tau }\mathcal{P}%
_{\nu \mu }^{\prime }\quad \mathrm{with}\quad \Lambda _{\nu \mu }=\nabla
_{\nu }\chi _{\mu }+\nabla _{\mu }\chi _{\nu }-{\frac{2}{3}}\delta _{\nu \mu
}\nabla _{\lambda }\chi _{\lambda },  \label{e3}
\end{equation}%
where $\tau $\ is the relaxation time due to interparticle collisions and $%
\chi _{\nu }\equiv \chi _{\nu }(\mathbf{r,}t)$ is the displacement field
which is related to the velocity field by $u_{\nu }(\mathbf{r,}t)=\partial
\chi _{\nu }(\mathbf{r,}t)/\partial t$. In the case of a spherical bubble,
the displacement field is given by \cite{lamb,blin}%
\begin{equation}
\chi _{\nu }(\mathbf{r,}t)=R^{3}r_{\nu }/3r^{3}\text{\qquad for\qquad }r\geq
R.  \label{chi1}
\end{equation}
We point out the quadrupole distortion of the Fermi surface does not distort
the spherical shape of the bubble if the displacement field $\chi _{\nu }(%
\mathbf{r,}t)$\ is given by Eq. (\ref{chi1}).

The external force $\mathbf{F}_{\mathrm{ext}}$\ in Eq. (\ref{e2}) is caused
by the vapor pressure on the liquid. The external force on the bubble
surface in the radially outward direction per unit area is given by \cite%
{ples49,bren95}
\begin{equation}
\overline{F}_{\mathrm{ext}}=P_{\mathrm{vap}}-P_{0},  \label{fext}
\end{equation}%
where $P_{\mathrm{vap}}$\ is the vapor pressure within the bubble and $P_{0}$%
\ is the pressure of the saturated vapor\ with respect to a plane surface.
Note that the condition of equilibrium between the bubble vapor and the
liquid is expressed by $P_{0}=P_{\mathrm{vap}}$ and $\overline{F}_{\mathrm{%
ext}}=0$, see Ref. \cite{frenk}.\

The Euler-like equations (\ref{e2}) can be reduced to the equation of motion
for the radius $R(t)$ of the bubble. Multiplying Eq. (\ref{e2}) by $\overset{%
\_}{\chi }_{\nu }(\mathbf{r,}t)=R^{2}r_{\nu }/r^{3}$, summing over $\nu $,
integrating over $\mathbf{r}$-space and using Eqs. (\ref{e3}), we obtain the
non-Markovian equation for the variable\ $R(t)$

\begin{equation}
B\overset{..}{R}+\frac{1}{2}\frac{\partial B}{\partial R}\overset{.}{R}%
^{2}+I(R;t)=-\frac{\partial \Phi }{\partial R}-4\pi R^{2}P_{0}\ (1-P_{%
\mathrm{vap}}/P_{0}),  \label{mem1}
\end{equation}%
where
\begin{equation}
I(R;t)=B\int_{t_{0}}^{t}dt^{\prime }\overset{.}{R}(t^{\prime })\ \exp
[(t^{\prime }-t)/\tau ]\ \mathcal{K}(t,t^{\prime })  \label{mint}
\end{equation}%
is the memory integral. The inertial parameter $B\equiv B(R)$\ and the
memory kernel $\mathcal{K}(t,t^{\prime })$ in Eqs. (\ref{mem1}) and (\ref%
{mint}) are given by \cite{lamb,blin,kosa03}
\begin{equation}
B(R)=4\pi m\rho _{0}R^{3},\quad \mathcal{K}(t,t^{\prime })=\frac{8\ \epsilon
_{F}}{5m\ R(t)R(t^{\prime })},  \label{b}
\end{equation}%
where $\rho _{0}$ is the bulk density of the nuclear matter and $\epsilon
_{F}$\ is the Fermi energy. Denoting the value of thermodynamic potential $%
\Phi $ corresponding to the absence of the bubble by $\Phi _{0}$, we have \
\begin{equation}
\Phi \equiv \Phi (R)=\Phi _{0}+\Delta \Phi (R),  \label{e1}
\end{equation}%
where $\Delta \Phi (R)$\ is the change of the thermodynamic potential due to
the variation of the bubble radius.\ The value of $\Delta \Phi (R)$ is given
by \cite{frenk}
\begin{equation}
\Delta \Phi (R)=4\pi \sigma \left( R^{2}-\frac{2R^{3}}{3R^{\ast }}\right) .
\label{phi5}
\end{equation}%
The thermodynamic potential $\Phi (R)$ for $R=R^{\ast }$ does not have a
minimum value, as in the case of equilibrium, but a maximum value, see
\textrm{Fig. 1}. The thermodynamic potential of Eq. (\ref{phi5}) corresponds
to the metastable liquid phase with $\mu _{\mathrm{liq}}>\mu _{\mathrm{vap}}$%
, where $\mu _{\mathrm{liq}}$ and $\mu _{\mathrm{vap}}$ are the chemical
potentials for the liquid and the bubble vapor, respectively.

\begin{figure}[tbp]
\includegraphics[width=4.0 in,height=4.0 in]{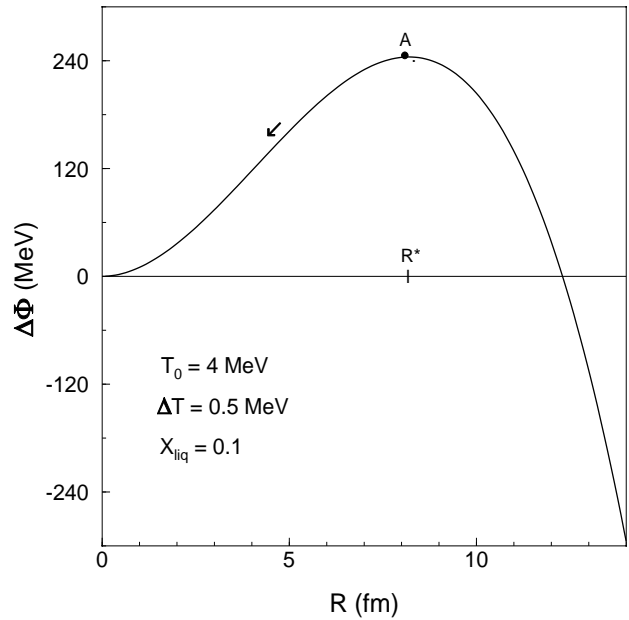}
\caption{ Dependence of the thermodynamical potential $\Delta \Phi
$\ (see Eq. (\ref{phi5})) of metastable (overheated) asymmetric
nuclear matter on the radius of the bubble. }
\end{figure}

Note that in
the Markovian limit at $\tau \rightarrow 0$\ the equation of motion (\ref%
{mem1}) is transformed to the generalized Rayleigh-Plesset equation of
classical bubble dynamics \cite{bren95}.

The critical radius $R^{\ast }$\ in Eq. (\ref{phi5}) depends on the
temperature $T_{0}$ of liquid-gas phase transition, the overheating
temperature $\Delta T$ and the isotopic asymmetry parameter $X_{\mathrm{liq}%
} $ ($X=(\rho _{n}-\rho _{p})/(\rho _{n}+\rho _{p})$, where $\rho _{n}$ and $%
\rho _{p}$ are the neutron and proton densities, respectively), see Eq. (\ref%
{rcrit}) and Refs. \cite{frenk,kosa03}. To evaluate the critical radius $%
R^{\ast }$ we point out that the caloric curve has the plateau region at the
temperature $T_{0}$ \cite{kssf}. Below we will adopt $T_{0}=4$ \textrm{MeV}
and $X_{\mathrm{liq}}=0.1$. If one assumes the process of isobaric heating
for the description of the plateau region in the caloric curve at $T_{0}=4$
\textrm{MeV}, the order of magnitude of the pressure $P_{0}$ should be $P_{%
\mathrm{0}}\approx 10^{-3}$~\textrm{MeV/fm}$^{3}$ for this process \cite%
{kssf}. Under these conditions, one obtains from Eq. (\ref{rcrit}) $R^{\ast
}=q/\Delta T$ with $q\approx 4.1\ \mathrm{MeV\cdot fm}$ for a symmetric
nuclear matter. The parameter $q$\ grows slightly with the asymmetry of the
nuclear matter.

In the case of boiling of asymmetric nuclear matter, the vapor asymmetry $X_{%
\mathrm{vap}}$ significantly exceeds the corresponding liquid asymmetry $X_{%
\mathrm{liq}}$ (at $X_{\mathrm{liq}}>0$), see Ref. \cite{kssf}. This is a
feature of the nuclear matter where the structure of the isospin symmetry
energy provides the condition $|\mu _{n}|<|\mu _{p}|$ and the preferable
evaporation of neutrons. Using this fact, we will neglect the proton
fraction in the bubble vapor. The presence of the noncondensable neutron
vapor within the bubble then gives a contribution to the term $\sim (1-P_{%
\mathrm{vap}}/P_{0})$ in Eq. (\ref{mem1}). (Note that the analogous term in
the classical Rayleigh-Plesset equation is due to the assumption that the
bubble contains some quantity of contaminant noncondensable gas \cite{bren95}%
.) The vapor pressure in the bubble includes both quantum and thermal
contributions.\ Since the temperature region of interest is $T\ll \epsilon
_{F}$, the thermal pressure is relatively small and for the degenerate
neutron gas one can use the Thomas-Fermi approximation with $P_{\mathrm{vap}%
}/P_{0}=(R^{\ast }/R)^{5}$ in Eq. (\ref{mem1}).

The surface tension coefficient $\sigma $\ in Eq. (\ref{phi5}) is
temperature dependent. We will use the following expression\ for $\sigma
\equiv \sigma (T)$\ \cite{rpl}
\begin{equation}
\sigma (T)=\sigma (0)\left[ ({T_{\mathrm{crit}}^{2}-T^{2}})/({T_{\mathrm{crit%
}}^{2}+T^{2}})\right] ^{5/4}\,\mathrm{,}  \label{sigma}
\end{equation}
where $T_{\mathrm{crit}}=14.6\,\mathrm{MeV}$ is the critical temperature for
infinite nuclear Fermi-liquid, associated with the SKM interaction \cite%
{kssf}.

3. Below we will consider the collapse phase, i.e. the descent from the top
of the barrier in Fig. 1 toward $R<R^{\ast }$. In general, an accurate
evaluation of the bubble dynamics requires to include a thermal boundary
condition at the bubble wall to provide the possible changes of the pressure
$P_{\mathrm{vap}}$ of the bubble vapor and its temperature \cite{bren95}. We
will neglect the thermal effects, assuming that the main contribution to the
pressure of the bubble vapor is due to the quantum motion of nucleons
because of $T\ll \epsilon _{F}$. If the collapse process is fast enough and
the condensation of the neutron's vapor within the bubble is negligible, the
force $\overline{F}_{\mathrm{ext}}$ of Eq. (\ref{fext}) has then to be taken
into consideration in Eq. (\ref{e2}). In an opposite case of a slow collapse
process, the possible growth of the pressure $P_{\mathrm{vap}}$\ at the
decrease of the bubble radius $R(t)$ is compensated by the condensation of
the vapor providing that $P_{\mathrm{vap}}=P_{0}=\mathrm{const}$ and $%
\overline{F}_{\mathrm{ext}}=0$.

The non-Markovian equation (\ref{mem1}) can be solved if one rewrites it as
a set of two connected equations for the bubble radius $R(t)$\ and the
memory integral $I(R;t)$ with the relevant initial conditions. The bubble
starts to collapse from the metastable state (point \textrm{A} on the top of
barrier in\textrm{\ Fig. 1}) at $R(0)=R^{\ast }$. Taking the overheating
temperature $\Delta T=0.5$~\textrm{MeV, }we obtain from Eq. (\ref{rcrit}) $%
R(0)=R^{\ast }=8.2$ \textrm{fm}. This critical radius $R^{\ast }=8.2$
\textrm{fm} seems too large for the finite nuclei. We point out, however,
that the inclusion of the Coulomb forces, which is ignored in our
consideration of the infinite nuclear matter, decreases the value of $%
R^{\ast }$\ \cite{blin86,blin}.\ Note, however, that our consideration
cannot be extended to very small values of $R^{\ast }$\ because for a small
enough bubble the finite size effects, in particular the finite diffuse
layer for the vapor density inside the bubble, can be important \cite{blin}%
.\ Using the above mentioned initial condition for \ $R(0)$, we restrict
ourselves to an infinite nuclear matter and expect that the main conclusions
concerning the Fermi motion effects on the collapse of the bubble in hot
Fermi liquid can be also applied to finite nuclei.

The initial velocity $\overset{.}{R}(0)$\ can be derived using the initial
kinetic energy $E_{\mathrm{kin,0}}$. Assuming the equipartition of energy
over degrees of freedom at $R=R^{\ast }$, we use $E_{\mathrm{kin,0}%
}=(T_{0}+\Delta T)/2$ and obtain $\overset{.}{R}(0)=-\sqrt{2E_{\mathrm{kin,0}%
}/B(R^{\ast })}=-2.9\cdot 10^{-3}c$, where sign \textquotedblright $-$%
\textquotedblright\ is used because of the collapse process with $%
R(t)<R^{\ast }$.

In \textrm{Fig. 2}, we show the time dependence of the radius $R(t)$ in the
presence of the noncondensable vapor of neutrons with $\overline{F}_{\mathrm{%
ext}}\neq 0$ (solid lines). Here we have adopted $\epsilon _{F}=37\ \mathrm{%
MeV,}$ $\sigma (0)=1.1\ \mathrm{MeV\cdot fm}^{-2}$ and\ $\rho _{0}=0.16\
\mathrm{fm}^{-3} $. For the relaxation time $\tau $\ we have used $\tau
=\tau _{0}/T^{2} $ with $\tau _{0}=850$ $\mathrm{fm}/c$ \cite{dani84}.\ As
seen from \textrm{Fig. 2}, the collapse of the bubble is accompanied by
oscillatory behavior of its radius $R(t)$.\ These oscillations occur due to
the memory integral in Eq. (\ref{mem1}) and disappear in the limit of
Markovian motion (see dashed line in Fig. 2). As seen from Fig. 2, the
collapse process stops at a certain values of the bubble radius, $R_{\min
}=0.28\ R^{\ast }$. The\ presence of non-zero $R_{\min }\neq 0$ is due to
the pressure of the noncondensable neutron vapor in the bubble. The value of
$R_{\min }$ is derived by the condition of the compensation of the adiabatic
forces in Eq. (\ref{mem1}) at $t\gg \tau $.

\begin{figure}[tbp]
\includegraphics[width=4.0 in,height=4.0 in]{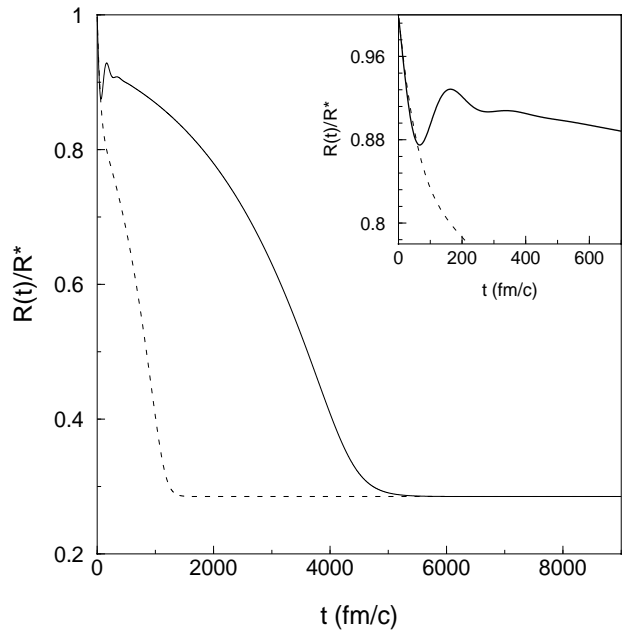}
\caption{Dependence of the radius $R(t)$\ of the collapsing bubble
on the time obtained for two regimes: non-Markovian motion, Eq.
(\ref{mem1}), (solid line) and Markovian (no memory) motion, Eq.
(\ref{nmem1}), with the friction coefficient $\protect\gamma $\
from Eq. (\ref{gamma2}) (dashed line). The inset shows the
collapse process near the top of barrier of $\Phi (R)$. The values
of $T_{0},\ \Delta T$ and $X_{{\rm liq}}$ as in Fig. 1. }
\end{figure}

The memory integral in Eq. (\ref%
{mem1}) leads to a significant delay in the collapse process. To show this
effect we will derive the friction coefficient $\gamma $ from the memory
integral $I(R;t)$. In a immediate vicinity of the critical radius $R^{\ast }$%
, the solution to Eq. (\ref{mem1}) takes the form of a superposition of
exponential functions $\exp (\lambda _{i}t)$ ($i=1,2$ and $3$) with
eigenvalues $\lambda _{i}$\ obtained as solutions to the secular equation
\begin{equation}
(\lambda ^{2}+k/B^{\ast })(\lambda +1/\tau )+(\widetilde{\kappa }/B^{\ast
})\lambda =0.  \label{disp4}
\end{equation}
Here $\ k=-4\pi (2\sigma -5P_{0}R^{\ast }),\mathrm{\quad }\widetilde{\kappa }%
=(32/5)\pi \rho _{0}\epsilon _{F}R^{\ast }$ and the mass coefficient $%
B^{\ast }\equiv B(R^{\ast })$ is taken from Eq. (\ref{b}) at $R=R^{\ast }$.
In the case of the zero-relaxation-time limit, $\tau \rightarrow 0,$ one
obtains from Eq. (\ref{disp4}) the motion with $\lambda =\pm \sqrt{%
|k|/B^{\ast }}$, i.e., the time evolution is derived by the static stiffness
coefficients $k$. In the opposite case of rare collisions, $\tau \rightarrow
\infty ,$ the solution to Eq. (\ref{disp4}) leads to a motion with $\lambda
=\pm i\sqrt{(-|k|+\widetilde{\kappa })/B^{\ast }}$, where the additional
contribution to the stiffness coefficient, $\widetilde{\kappa }$, appears
because of Fermi surface distortion effect. In both limits $\tau \rightarrow
0$ and $\tau \rightarrow \infty ,$\ the macroscopic equation of motion (\ref%
{mem1}) is reduced to the usual Markovian (no memory) equation with a
friction coefficient $\gamma $%
\begin{equation}
m\rho _{0}R\overset{..}{R}+\frac{3}{2}m\rho _{0}\overset{.}{R}^{2}+\gamma
\overset{.}{R}=-\frac{2\sigma }{R}\left( 1-\frac{R}{R^{\ast }}\right)
-P_{0}\ \left( 1-\left( \frac{R^{\ast }}{R}\right) ^{5}\right) .
\label{nmem1}
\end{equation}
The friction coefficient $\gamma $\ is related to the relaxation time $\tau $%
\ as \cite{kora01}

\begin{equation}
\gamma =\omega _{F}\ B^{\ast }\ \frac{\omega _{F}\ \tau }{1+(\omega _{F}\
\tau )^{2}},  \label{gamma2}
\end{equation}
where $\omega _{F}=\sqrt{\widetilde{\kappa }/B^{\ast }}$ is the
characteristic frequency for the eigenvibrations caused by the Fermi surface
distortion. The corresponding solution for $R(t)$ obtained from the
Markovian equation of motion (\ref{nmem1}) with friction coefficient $\gamma
$, from Eq. (\ref{gamma2}), is shown in \textrm{Fig. 2} by the dashed lines.
As seen from Fig. 2 the presence of memory effect (solid line) strongly
hinders the collapse process. If the collapse process becomes slow enough
the growth of the vapor pressure $P_{\mathrm{vap}}$\ is compensated by the
condensation of the vapor and the contribution from $\overline{F}_{\mathrm{%
ext}} $\ is negligible in Eq. (\ref{mem1}). The collapse process then leads
to the disappearance of the bubble.

In \textrm{Fig. 3} we have plotted the dependence of the time, $t_{\mathrm{%
clps}}$, required for the total collapse from $R=R^{\ast }$ to $R=0$, on the
relaxation time $\tau $, for the bubble in the absence of the force $%
\overline{F}_{\mathrm{ext}}$ (i.e., neglecting the term $\sim P_{0}$\ in
Eqs. (\ref{mem1}) and (\ref{nmem1})).

In \textrm{Fig. 3}\ we compare the results for $t_{\mathrm{clps}}$\ for both
the non-Markovian motion derived by Eq. (\ref{mem1}) (solid line) and the
usual Markovian (no memory) one given by Eq. (\ref{nmem1})\ (dashed line).
As seen from \textrm{Fig. 3}, the behavior of $t_{\mathrm{clps}}$ is changed
significantly due to the memory effects. A delay in the collapse of the
bubble occurs due the non-Markovian (memory) effect for the large values of
the relaxation time.

\begin{figure}[tbp]
\includegraphics[width=4.0 in,height=4.0 in]{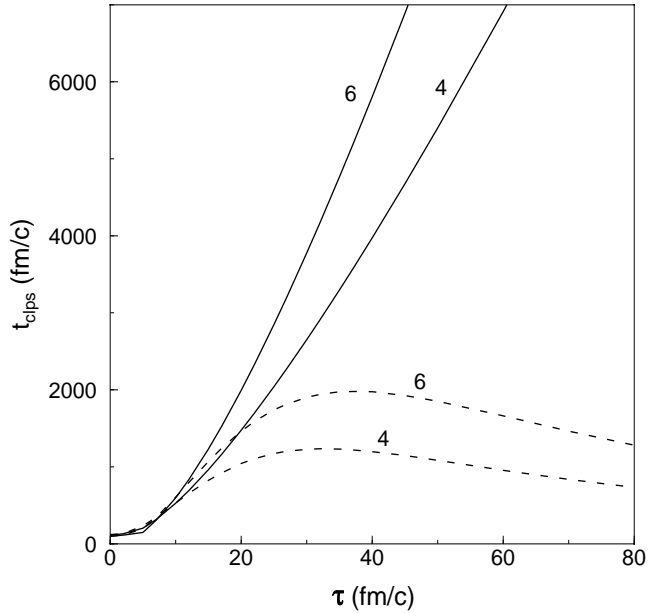}
\caption{The collapse time $t_{{\rm clps}}$\ for the metastable
bubble in an overheated Fermi liquid as a function of relaxation
time $\protect\tau $ for different temperatures $T_0=4$ and 6 MeV
shown near the curves at fixed asymmetry parameter $X_{{\rm
liq}}=0.1$. The solid and dashed lines represent the calculations
with non-Markovian and Markovian equations of motion respectively.
The values of $\Delta T$ and $X_{{\rm liq}}$ are the same as in
Fig. 1. }
\end{figure}

In the limit of a non-viscous liquid, $\tau \rightarrow 0$, we
obtain for both motions\ $t_{\mathrm{clps,0}}=4\cdot 10^{-22}$
\textrm{s}. This result can be compared with the classical one of Rayleigh $%
t_{\mathrm{clps,R}}$ \cite{lord}. To do that, we note that under the
condition used above for $T_{0}=4\ \mathrm{MeV},$ $P_{\mathrm{0}}\approx
10^{-3}~\mathrm{MeV/fm}^{3},\ X_{\mathrm{liq}}=0.1$\ and $R^{\ast }=8.2$
\textrm{fm}, the Fermi liquid is undercompressed by a pressure $\Delta
P\approx 0.25\ \mathrm{MeV/fm}^{3}$. The Rayleigh collapsing time $t_{%
\mathrm{clps,R}}$ is given by \cite{bren95}
\begin{equation*}
t_{\mathrm{clps,R}}=0.915\left( \frac{m\rho _{\mathrm{liq}}R^{2}(0)}{\Delta P%
}\right) ^{1/2}
\end{equation*}
with the yield $t_{\mathrm{clps,R}}=6.1\cdot 10^{-22}$ \textrm{s }which is%
\textrm{\ }in a good agreement with $t_{\mathrm{clps,0}}$.

4. In conclusion we note that the collapse of the bubble in an overheated
(undercompressed) Fermi liquid is strongly influenced by memory effects, if
the relaxation time $\tau $\ is large enough at $\omega _{F}\ \tau \gtrsim 1$%
. In this case, the collapse of the bubble is accompanied by characteristic
shape oscillations of\ the bubble radius $R(t)$\ (see Fig. 2) which depend
on the memory kernel $\mathcal{K}(t,t^{\prime })$ and the relaxation time $%
\tau $. These oscillations appear due to the non-adiabatic elastic force
induced by the memory integral. The non-adiabatic elastic force\ acts
against the adiabatic force $-\partial \Phi /\partial R$\ (see Eq. (\ref%
{mem1})) and hinders the collapse of the bubble. In contrast to the case of
the Markovian motion, the delay in the collapse is caused here by both
conservative elastic and friction forces.

We should like to point out that the study of the cavitation in a
two-component nuclear matter could represent a stimulating area of interest
due to the fact that the vapor inside embryonic bubbles is strongly
asymmetric system with $(N-Z)/(N+Z)\approx 1,$ where $N$\ and $Z$\ are the
number of neutrons and protons, respectively. The collapse of such extremely
neutron rich system in which the energy is focused in a small region could
show up new phenomena in the clusterization of nucleons.

The main goal of this paper was to show the existence of a new phenomena of
shape oscillations of\ the collapsing bubbles in a hot nuclear Fermi liquid.
This phenomena does not exist in classical (Newtonian) liquids.\ The shape
oscillations of the collapsing bubble could induce a quite stimulating
search of the accompanied characteristic $\gamma $-quanta emission. That
provides, in principle, the possibility for the measurement of the
temperature of the first kind phase transition through the measurement of
the energy and the damping of the corresponding resonances in the $\gamma $%
-quanta spectrum.

Our consideration was restricted to a nuclear matter for maximum simplicity.
Both effects of Coulomb forces and nuclear surface tension must be taken
into account in the case of bubble dynamics in finite nuclei. The presence
of Coulomb forces enhances the bubble formation and decreases the critical
radius $R^{\ast }$. Note, however, that the minimum value of the radius $%
R^{\ast }$,\ for which our approach is applicable, is determined by the
condition $a/R^{\ast }\ll 1$, where $a=0.5\div 1$ fm\ is the surface
thickness of the bubble \cite{blin}. The bubble dynamics in hot finite
nucleus requires further investigations.

\bigskip

The author thanks H. Hofmann and R. Hilton for valuable suggestions and the
Physics Department of the Technische Universit\"{a}t M\"{u}nchen for the
nice hospitality. Financial support by the Deutsche Forschungsgemeinschaft
under contract 436 UKR 113/66/0-1 is gratefully acknowledged.

\end{document}